\documentclass[pra,reprint,superscriptaddress,amsmath,amssymb]{revtex4-2}
\usepackage{graphicx}
\usepackage[colorlinks=true,linkcolor=blue,citecolor=blue,urlcolor=blue]{hyperref}

\begin{document}

\title{Self-guided estimation of the stabilizer fidelity with single-copy Pauli measurements}

\author{M.\,Torres}
\affiliation{Dipartimento di Fisica ``E. Pancini'', Universit\`a di Napoli Federico II - INFN sezione di Napoli, Complesso Universitario di Monte S. Angelo Edificio 6, via Cintia, 80126 Napoli, Italy}

\author{J.\,Gidi}
\affiliation{Instituto de F\'{\i}sica Fundamental (IFF-CSIC), Calle Serrano 113b, Madrid 28006, Spain}
\affiliation{Quantum Advanced Research Center (QuARC-CSIC), Calle Serrano 113b, Madrid 28006, Spain}

\author{L.\,Pereira}
\affiliation{ICFO - Institut de Ci\`encies Fot\`oniques, The Barcelona Institute of Science and Technology, 08860 Castelldefels, Barcelona, Spain}

\date{\today}

\begin{abstract}
The stabilizer fidelity quantifies the non-stabilizerness of pure states, but it is hard to estimate because it requires optimizing over $2^{O(n^2)}$ stabilizer states for $n$ qubits. Here, we propose a self-guided estimation protocol for the stabilizer fidelity of an unknown state, pure or mixed, which requires only Pauli measurements on single copies. The protocol moves through the graph of neighboring stabilizer states, obtains the fidelities of a whole stabilizer basis from a single direct fidelity estimation, and reuses previous outcomes, which also provide a restart point that mitigates the barren plateau of a random start. In addition, it certifies the global optimum when the lower confidence bound of the estimated fidelity exceeds $\cos^2(\pi/8)$. In numerical simulations of up to 14 qubits, the protocol reached the optimum in all runs, with the number of iterations remaining small and increasing only slightly with $n$.
\end{abstract}

\maketitle

\section{Introduction}

The resources that make quantum computers more powerful than classical ones can be quantified through their distance to efficiently simulable states. In particular, Clifford circuits acting on stabilizer states can be simulated in polynomial time~\cite{Aaronson2004}, so that the non-stabilizerness of a state, also known as magic, is a necessary resource for quantum advantage~\cite{BravyiKitaev2005,Veitch2014,Howard2017}. A natural measure of non-stabilizerness is the stabilizer fidelity $F_{\rm STAB}$, that is, the largest fidelity between the state and a stabilizer state~\cite{Bravyi2019}. Its estimation from experimental data allows certifying how close the states prepared in current quantum devices are to a stabilizer state and, for pure states, when $F_{\rm STAB}<1$ is certified, their non-stabilizerness~\cite{Wagner2025,Haug2024,Kliesch2021,Eisert2020}.

Estimating $F_{\rm STAB}$ is hard, since the number of stabilizer states grows as $2^{O(n^2)}$ with the number $n$ of qubits. Moreover, the existence of pseudomagic states~\cite{Gu2024} implies that no efficient method can estimate $F_{\rm STAB}$ in all regimes~\cite{Chen2024}. Even when the state vector is known, $F_{\rm STAB}$ has been computed only up to 9 qubits for Haar-random states~\cite{Hamaguchi2024}. For unknown pure states, the stabilizer R\'enyi entropies~\cite{Leone2022} provide only upper bounds on $F_{\rm STAB}$~\cite{Haug2023}. Beyond such bounds, most approaches rely on joint measurements on two copies of the state. For pure states, Bell difference sampling~\cite{Montanaro2017,Gross2021,Hangleiter2023} finds a stabilizer state whose fidelity is close to $F_{\rm STAB}$~\cite{Grewal2023}. For mixed states, two-copy measurements find the closest stabilizer product state~\cite{Grewal2026} and, through stabilizer bootstrapping, estimate $F_{\rm STAB}$~\cite{Chen2024}. Two-copy measurements also enable tolerant testers that decide whether $F_{\rm STAB}$ is high or low~\cite{Arunachalam2024,Bao2024,Iyer2024}, as well as the learning of $t$-doped states via Bell measurements~\cite{LaiCheng2022,Leone2024learn}. Single-copy measurements, in contrast, have been explored mainly for testing and learning. Random Clifford measurements decide whether a state is a stabilizer state~\cite{Hinsche2025}, adaptive Clifford measurements find a near-optimal stabilizer state above a fidelity threshold, even for mixed states~\cite{Bittel2026}, and local Clifford circuits of logarithmic depth learn stabilizer groups on average~\cite{ChoKim2026}. Likewise, $t$-doped states can be learned from single copies~\cite{Chia2024,Grewal2025}. These methods, however, aim to identify or reconstruct a state rather than to estimate $F_{\rm STAB}$, and they rely on Clifford rather than Pauli measurements. Consequently, apart from an exhaustive search over stabilizer states, no current method directly estimates $F_{\rm STAB}$ using only Pauli measurements on single copies, the simplest measurements available in current devices.

A natural alternative is a variational approach~\cite{Peruzzo2014,Kandala2017,Cerezo2021VQA,Cerezo2020VQFE,Gidi2023}, where the fidelity with a candidate state is measured and optimized by a classical routine, as in self-guided quantum tomography~\cite{Ferrie2014,Chapman2016,Utreras2019,Hou2020,Zambrano2020,Rambach2021}. However, the fidelity is a global function, whose differences between candidates decrease exponentially with $n$ from a random start, which is known as the barren plateau phenomenon~\cite{McClean2018,Cerezo2021BP}. Local cost functions avoid barren plateaus in shallow circuits~\cite{Cerezo2021BP}, and two-qubit local functions provide a good starting point for the geometric measure of entanglement~\cite{MunozMoller2022,Zambrano2023}. Despite these advances, the set of stabilizer states is discrete, so that gradient-based optimization does not apply, and it is not clear how to search over it with single-copy measurements and how to mitigate its barren plateau.

In this article, we propose a self-guided estimation protocol for the stabilizer fidelity of an unknown state, pure or mixed, which requires only Pauli measurements on single copies. The protocol keeps a stabilizer state $|\phi\rangle$ as the current estimate and moves to the best of a few candidates, mostly among its $4(2^n-1)$ neighbors, that is, the stabilizer states with overlap $1/2$ with $|\phi\rangle$~\cite{Garcia2017}, whose fidelities are estimated by direct fidelity estimation (DFE)~\cite{Flammia2011,daSilva2011,Pallister2018,ZhuHayashi2019,BarberaRodriguez2025,ChaLee2025}. A single DFE run on the group of $|\phi\rangle$ provides the fidelity of any state of its stabilizer basis, in analogy with the error rates of Pauli channels~\cite{Flammia2020}, and we show that only a few directions can contain a better neighbor, which a prefix search finds without enumerating the $2^n$ directions~\cite{Goldreich1989,Kushilevitz1993}, and that $|\phi\rangle$ and its neighbors share half of their Pauli strings. A Pauli store reuses these outcomes, and every measured string also provides outcomes for its substrings~\cite{Cotler2020,Huang2020}, so that the store contains the expectation values of the Pauli strings of few qubits, from which we build a restart point that mitigates the barren plateau. The classical cost of the protocol is polynomial in $n$ for circuits of constant depth, and the protocol certifies the global optimum when the lower confidence bound of its estimated fidelity exceeds $\cos^2(\pi/8)$.

We carried out numerical simulations of up to 14 qubits, starting always from a random stabilizer state, for four families of states whose stabilizer fidelity is known exactly, namely stabilizer states with a random perturbation, T-doped states prepared by Clifford circuits of depth $O(n)$ or $O(\log n)$, and depolarized stabilizer states. The protocol reached the optimum in all runs, without false certificates. The number of iterations remained small and increased only slightly with $n$, while the number of shots per iteration increased with $n$, as the reuse of the Pauli store decreases.

\section{Theoretical background}

In this section, we introduce the concepts of stabilizer fidelity and direct fidelity estimation, on which our protocol is built.

\subsection{Stabilizer states and stabilizer fidelity}

Let us consider $n$ qubits and the Pauli strings $P_\kappa=i^{a\cdot b}X^{a}Z^{b}$, labeled by $\kappa=(a,b)\in\mathbb{F}_2^{2n}$, with $a \cdot b$ computed over the integers, where $\mathbb{F}_2^{m}$ is the set of binary vectors of length $m$ with addition modulo 2. A stabilizer state $|\phi\rangle$ is the unique common eigenstate with eigenvalue $+1$ of an Abelian group $S$ of $2^n$ signed Pauli strings, which is generated by $n$ independent signed strings $s_ig_i$, where $g_i=P_{\kappa_i}$ is a Pauli string and $s_i=\pm1$ is its sign. The elements of $S$ are products of the generators, that is, $s_ug_u=\prod_i(s_ig_i)^{u_i}$ with $u\in\mathbb{F}_2^{n}$, where $s_u=\pm1$ and $g_u$ is a Pauli string, and their average over the group gives the projector onto $|\phi\rangle$
\begin{equation}
|\phi\rangle\langle \phi|=\frac{1}{2^n}\sum_{u\in\mathbb{F}_2^n}s_u\,g_u .
\label{eq:projector}
\end{equation}
The stabilizer fidelity of a state $\rho$, pure or mixed, is defined as the largest fidelity between $\rho$ and a stabilizer state~\cite{Bravyi2019},
\begin{equation}
F_{\rm STAB}(\rho)=\max_{|\eta\rangle\in{\rm Stab}_n}\langle \eta|\rho|\eta\rangle,
\label{eq:fstab}
\end{equation}
where ${\rm Stab}_n$ is the set of $n$-qubit stabilizer states, which contains $2^n\prod_{j=1}^n(2^j+1)=2^{O(n^2)}$ states~\cite{Aaronson2004, Chen2024}. Thereby, an exhaustive search over ${\rm Stab}_n$ is only feasible for a few qubits.

\subsection{Direct fidelity estimation}

DFE estimates the fidelity between an unknown state $\rho$ and a known pure state by sampling Pauli strings with probability proportional to the squared coefficients of the known state~\cite{Flammia2011,daSilva2011}. For a stabilizer target these coefficients are uniform on the group, so that Eq.\thinspace\eqref{eq:projector} leads to
\begin{equation}
f(\phi)=\langle \phi|\rho|\phi\rangle=\mathbb{E}_{u}\big[s_u\,\mathrm{tr}(\rho\,g_u)\big],
\label{eq:dfe}
\end{equation}
with $u$ uniform in $\mathbb{F}_2^n$. Then, the estimator measures $N$ strings $g_{u_j}$ once each on fresh copies of $\rho$, with outcomes $P_j=\pm1$, and returns $\hat f=\frac{1}{N}\sum_js_{u_j}P_j$. Since each term lies in $[-1,1]$, $N=2\ln(2/\delta)/\varepsilon^2$ shots give an error below $\varepsilon$ with probability at least $1-\delta$ for any $n$, in contrast with non-stabilizer targets, whose DFE becomes harder with their non-stabilizerness~\cite{Leone2023}.

\section{Self-guided estimation protocol}

In this section, we introduce the neighbors of a stabilizer state, and then we describe the search, the Pauli store, the local restart and the classical cost of the protocol.

\subsection{Stabilizer neighbors}
\label{sec:neighbors}

\begin{figure*}[t]
\includegraphics[width=\textwidth]{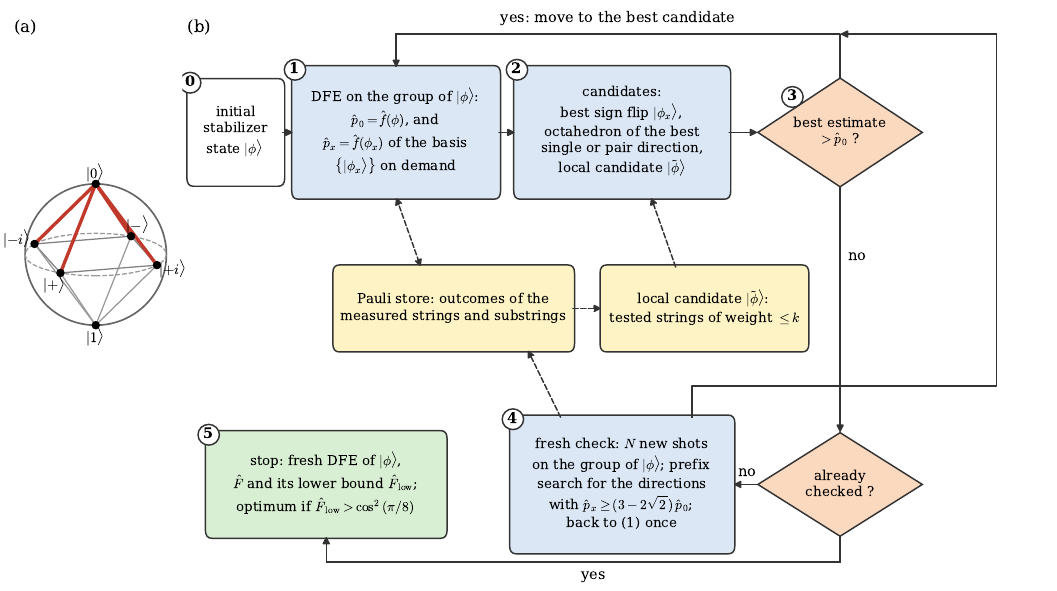}
\caption{Self-guided estimation protocol. Inset (a) shows the single-qubit stabilizer states on the Bloch sphere, where the neighbors of $|0\rangle$ (red edges) are the four states on the equator. Inset (b) shows the flow diagram of the protocol, whose steps are described in Sec.~\ref{sec:search}.}
\label{fig:scheme}
\end{figure*}

Let $|\phi\rangle$ be the current stabilizer state, with generators $s_ig_i$ and destabilizers $d_i$, that is, Pauli strings that anticommute with $g_i$ and commute with all other generators and destabilizers~\cite{Aaronson2004}. The product $D_x=\prod_id_i^{x_i}$, with $x\in\mathbb{F}_2^n$, flips the signs of the generators with $x_i=1$, so that the states $|\phi_x\rangle=D_x|\phi\rangle$ form a stabilizer basis with $|\phi_0\rangle=|\phi\rangle$. We define the neighbors of $|\phi\rangle$ as
\begin{equation}
|\nu_{x,m}\rangle=\frac{|\phi\rangle+i^m|\phi_x\rangle}{\sqrt{2}},\qquad x\neq0,\ m\in\{0,1,2,3\},
\label{eq:neighbors}
\end{equation}
which are the $4(2^n-1)$ stabilizer states with overlap $|\langle\phi|\nu_{x,m}\rangle|^2=1/2$, the largest overlap between two different stabilizer states~\cite{Garcia2017}. For one qubit, the neighbors of $|0\rangle$ are the four states on the equator of the Bloch sphere, as shown in Fig.\thinspace\ref{fig:scheme}(a), and, in general, $|\phi\rangle$, $|\phi_x\rangle$ and the four neighbors in direction $x$ form an octahedron.

The fidelities of the neighbors are determined by $p_x=\langle \phi_x|\rho|\phi_x\rangle$ and $r_x=\langle \phi|\rho|\phi_x\rangle$ as
\begin{equation}
f(\nu_{x,m})=\frac{p_0+p_x}{2}+{\rm Re}\big(i^mr_x\big),\qquad |r_x|\le\sqrt{p_0\,p_x},
\label{eq:neighborfid}
\end{equation}
for pure and mixed states (Appendix~\ref{app:neighbors}). Therefore, the best neighbor in direction $x$ has a fidelity of at most $(\sqrt{p_0}+\sqrt{p_x})^2/2$, which increases with $p_x$ and can exceed $f(\phi)=p_0$ only if $p_x>(3-2\sqrt2)\,p_0\approx0.17\,p_0$. Since the $p_x$ sum to one, at most $(1-p_0)/[(3-2\sqrt2)\,p_0]$ directions satisfy this condition, that is, only a few of the $2^n-1$ directions deserve to be explored.

The probabilities $p_x$ are obtained from the DFE of $|\phi\rangle$ without additional measurements, since Eq.\thinspace\eqref{eq:dfe} generalizes to $p_x=\mathbb{E}_u[(-1)^{x\cdot u}s_u\mathrm{tr}(\rho\,g_u)]$ (Appendix~\ref{app:walsh}). Each $\hat p_x$ is a sum over the measured group elements, so that the protocol evaluates only the values that it needs, namely those of the $n$ single directions $x=e_i$, of the $n(n-1)/2$ pair directions $x=e_i+e_j$, and of the best sign flip, together with those of the few directions above the threshold when no candidate improves, which a prefix search finds whatever their weight (Appendix~\ref{app:walsh}). In addition, the octahedron of direction $x$ uses only four cosets of the subgroup $T_x\subset S$ of strings that commute with $D_x$, and two of them belong to $S$ (Appendix~\ref{app:neighbors}). Thereby, the DFE of a neighbor shares half of its Pauli strings with the DFE of $|\phi\rangle$.

\subsection{Self-guided search}
\label{sec:search}

The protocol moves $|\phi\rangle$ through the graph of neighbors, as shown in Fig.\thinspace\ref{fig:scheme}(b), according to the following steps.
\begin{enumerate}
    \item[0.] Selection of a stabilizer state $|\phi\rangle$ to initialize the search.
    \item DFE on the group of $|\phi\rangle$ with at most $N$ new shots, which gives $\hat f(\phi)=\hat p_0$ and any $\hat p_x$ on demand, and whose outcomes are kept in the Pauli store.
    \item Estimation of three kinds of candidates, that is, (i) the sign flip $|\phi_x\rangle$ that maximizes $\hat p_x$, (ii) the octahedron of the single or pair direction $d$ with the largest $\hat p_d$, formed by the flip $|\phi_d\rangle$, whose estimate is $\hat p_d$, and the four neighbors $|\nu_{d,m}\rangle$, whose DFEs reuse the stored outcomes, and (iii) the local candidate $|\tilde\phi\rangle$ built from the stored Pauli strings of few qubits (Sec.~\ref{sec:restart}).
    \item Move to the best candidate if its estimate improves on $\hat p_0$, and return to step (1).
    \item Freshly check if there is no improvement. That is, $N$ new shots on uniformly drawn elements of the group of $|\phi\rangle$ are added to the store, and a prefix search finds the directions with $\hat p_x\ge(3-2\sqrt2)\,\hat p_0$ with the values of $\hat p_0$ and $\hat p_x$ of step (1). Then, steps (1)--(3) are repeated once with the octahedra of these directions, where $\hat p_0$ and the candidates are estimated with the updated store, while the flips of these directions keep the values $\hat p_x$ that selected them. The search stops if no candidate improves, or after a predefined maximum number of iterations.
    \item Fresh DFE of the output with $N_f$ shots, which gives $\hat F$ and its lower confidence bound $\hat F_{\rm low}=\hat F-\varepsilon$.
\end{enumerate}
We use $N=1000$ and $N_f=\lceil2\ln(1/\delta)/\varepsilon^2\rceil=10234$, with $\varepsilon=0.03$ and $\delta=0.01$, so that $f(\phi)\ge\hat F_{\rm low}$ with probability at least $1-\delta$ (Appendix~\ref{app:cert}).

The best sign flip is found by a coordinate ascent of $\hat p_x$ over the bits of $x$, started from $x=0$ and from the solution of the linear system $x\cdot u=[\hat c(u)<0]$ over $\mathbb{F}_2$ for $n$ independent measured elements $u$, where $c(u)=s_u\mathrm{tr}(\rho\,g_u)$. The second starting point is necessary, since an ascent from $x=0$ can miss the flips of several generators, while it is exact for exact signs whenever one basis state has $p_{x^*}>1/2$, since then every $c(u)$ has the sign $(-1)^{x^*\cdot u}$ (Appendix~\ref{app:walsh}).

The fresh check corrects two effects of the reuse of outcomes. On the one hand, every candidate is the best of many noisy estimates, scored with the outcomes that selected it, and the estimate of the current state can be inflated by its selection as the best move. On the other hand, this estimate no longer changes when all the elements of the current group are stored, since Eq.\thinspace\eqref{eq:store} then draws no new shots. In addition, the single and pair directions depend on the choice of generators, while the best direction can combine several of them, which the prefix search finds for exact values and $\hat{p}_0 \ge 0.16$, whatever its weight (Appendix~\ref{app:walsh}).

Like any local search, the protocol can stop at a local maximum of the graph of neighbors, such as $|+\rangle^{\otimes n}$ for $\rho=0.6\,|0\rangle\!\langle0|^{\otimes n}+0.4\,|+\rangle\!\langle+|^{\otimes n}$ with $n\ge2$. Since the output is a stabilizer state, $\hat F$ is a lower estimate of $F_{\rm STAB}$. However, at most one stabilizer state has a fidelity above $\cos^2(\pi/8)$, since two different stabilizer states satisfy $\langle \eta_1|\rho|\eta_1\rangle+\langle \eta_2|\rho|\eta_2\rangle\le1+1/\sqrt2$ for any $\rho$~\cite{Garcia2017,Grewal2023} (Appendix~\ref{app:cert}). Therefore, if $\hat F_{\rm low}>\cos^2(\pi/8)\approx0.854$, the output is the unique global optimum and $\hat F$ estimates $F_{\rm STAB}$, while below this threshold it certifies a lower bound on $F_{\rm STAB}$.

\subsection{Pauli store}
\label{sec:store}

Consecutive candidates share many Pauli strings, as shown in Sec.~\ref{sec:neighbors}. In order to reuse their outcomes, the Pauli store keeps, for every measured unsigned Pauli string $P$ and for its substrings of weight at most $k$, the number $m_P$ of outcomes and their mean $\bar P$. Then, the DFE of a target with group $U$ splits it into the reused set $K\subseteq U$ of strings with stored outcomes, each with weight $1/|U|$, and the sampled set $R=U\setminus K$, which receives $N_R=\lceil N|R|/|U|\rceil$ new shots on uniformly drawn strings. The estimate is
\begin{equation}
\hat A_U=\frac{1}{|U|}\sum_{P\in K}s_P\,\bar P+\frac{|R|}{|U|}\,\frac{1}{N_R}\sum_{j=1}^{N_R}s_jP_j,
\label{eq:store}
\end{equation}
where $s_j$ and $P_j=\pm1$ are the sign and the outcome of the $j$-th sampled string. For queries fixed in advance, this estimator is unbiased (Appendix~\ref{app:store}), although its variance can exceed that of a new DFE with $N$ shots when $|U|<N$ and many strings of $K$ have fewer than $N/|U|$ outcomes, since each of them keeps the weight $1/|U|$. Since the search queries depend on previous outcomes, the store is used only to decide where to move, and the reported fidelity is always estimated using fresh copies.

The reuse is limited by the number of elements stored in $U$. After the DFE of $|\phi\rangle$, a neighbor finds stored essentially only the sampled elements of $T_x$, so that $|K|/|U|\approx[1-(1-2^{-n})^N]/2$, which is close to $1/2$ for $2^n\ll N$ and decreases as $N/2^{n+1}$ for $2^n\gg N$. Therefore, the store saves shots mainly for few qubits. In addition, the set $K$ is found by decomposing the stored strings in the group, at a cost linear in their number, so that the elements of $U$ are never enumerated.

The store also exploits the structure of a single measurement. Measuring a string $P$ in its product eigenbasis gives one bit per qubit, and the parity of these bits on the support of a substring $P'$, that is, $P$ with some letters replaced by the identity, is an unbiased outcome of $P'$. In this way, every shot also feeds the substrings of weight at most $k$ in the measured string, increasing reuse between candidates and yielding the few-qubit expectation values of the local restart.

\subsection{Local restart}
\label{sec:restart}

From a random stabilizer state, the fidelities of all neighbors are on the order of $2^{-n}$, so their differences are smaller than the DFE noise $1/\sqrt{N}$, which is a discrete analog of the barren plateau phenomenon~\cite{McClean2018,Cerezo2021BP}. Moreover, the neighbors of step (2) lie only along single and pair directions, in order to keep the classical cost polynomial, so that from a random start the search can stop before the region of high fidelity. In fact, without the local restart, the moves in step (2) stopped before this region, even with exact fidelity, whereas with $N$ shots the search reached it only after a long random walk driven by statistical noise (Appendix~\ref{app:sims}).

In order to mitigate this barren plateau, we build an additional candidate from the few-qubit data of the store, which does not rely on the global fidelity. If the optimal stabilizer state has generators of weight at most $k$, these generators have large expectation values $|\langle P\rangle|$, and so do their estimates $|\bar P|$. Then, the local restart builds a candidate in three steps.
\begin{enumerate}
\item Significance test. For every stored string $P$ of weight at most $k$, we compute the Hoeffding $p$-value $p_P=2e^{-m_P\bar P^2/2}$ of the hypothesis $\langle P\rangle=0$, and we call $P$ significant if $p_P\le\delta/M$, where $M=\sum_{w=1}^{k}\binom{n}{w}3^w$ is the number of non-identity Pauli strings of weight at most $k$ and $\delta$ is that of the certificate.
\item Ranking. The significant strings come first, in decreasing order of $|\bar P|$, followed by the other strings in increasing order of $p_P$, since for them $|\bar P|$ is not reliable.
\item Greedy selection. Going down the ranking, a string is accepted with the sign of $\bar P$ if it commutes with the accepted strings and is not a product of them, until $n$ generators are accepted.
\end{enumerate}
The candidate $|\tilde\phi\rangle$ defined by these generators is built at every iteration and compared in step (3) through its DFE. Appendix~\ref{app:example} shows an example for four qubits.

The significance test is necessary, since most stored strings have few outcomes and many of them have $|\bar P|=1$ by chance. This test is the Bonferroni correction~\cite{Dunn1961}, for which $M$ is fixed before any measurement, so that, for numbers of outcomes fixed in advance, no string with $\langle P\rangle=0$ is selected with probability at least $1-\delta$, whatever the correlations between strings that share shots. In the search the numbers of outcomes depend on previous outcomes, but an error of the ranking can only lead to a poor move, since the reported fidelity is always estimated with fresh copies, although the DFE of the candidate can reuse the outcomes of the ranking. For the same reason, the candidate is always completed to $n$ generators, if necessary with strings that are not significant or with further commuting strings with sign $+1$, since a poor candidate is usually discarded by its DFE.

The local restart maximizes greedily the local cost
\begin{equation}
C_L(s_1g_1,\dots,s_ng_n)=\frac12+\frac{1}{2n}\sum_{i=1}^ns_i\langle g_i\rangle ,
\label{eq:localcost}
\end{equation}
over the stabilizer states with generators of weight at most $k$, where $C_L$ depends on the generators and not only on the state that they define. Each term of Eq.\thinspace\eqref{eq:localcost} changes by order $1/n$ when one generator is found, while the fidelity changes by order $2^{-n}$ far from the optimum. In addition, $1-n\,(1-C_L)\le f(\phi)\le C_L$, since $|\phi\rangle\langle\phi|$ is the product of the projectors $(I+s_ig_i)/2$, as for the local and global costs of Ref.\,\cite{Cerezo2021BP}, so that a state with $1-C_L\ll1/n$ also has a fidelity close to one. Thereby, maximizing $C_L$ leads the search to the region of high fidelity, where the moves of step (2) are effective. For noisy states this lower bound is not informative, and the local restart relies on the generators of the optimal state having the largest expectation values, as for the depolarized stabilizer states of Sec.~\ref{sec:setup}.

The local restart requires that the optimal stabilizer state has generators of weight at most $k$, which holds for states prepared by Clifford circuits whose light cone covers at most $k$ qubits. A circuit of depth $d$ with two-qubit gates between neighboring qubits has a light cone of at most $2d$ qubits, so that we use $k=\max(4,2\lceil\log_2n\rceil)$, which covers $\lceil\log_2n\rceil$ layers of such gates, while $k = 2d$ is enough for circuits of known constant depth $d$. Notice that $k\ge n$ for $n\le6$, so that in this range the store keeps every substring of the measured strings, and the restart is local only for larger $n$.

\subsection{Classical cost}
\label{sec:cost}

A shot of a string of weight $w$ feeds $\binom{w}{\le k}=\sum_{j\le k}\binom{w}{j}$ substrings, and an iteration decomposes the $M_{\rm s}$ stored strings in the groups of the candidates, merges the new outcomes into the store, and evaluates the $\hat p_x$ of the single and pair directions on the $L\le M_{\rm s}+N$ group elements with data. In addition, the prefix search keeps at most $B\le64$ prefixes at each of its $n$ levels. Therefore, a search with $S$ shots and $I$ iterations costs
\begin{equation}
O\Big(S\tbinom{w}{\le k}+I\big(nM_{\rm s}+M_{\rm s}\log M_{\rm s}+n^2L+nBL\big)\Big)
\label{eq:cost}
\end{equation}
operations, with $M_{\rm s}\le S\binom{w}{\le k}$, where a product of two Pauli strings counts as one operation, as for $n\le64$ with strings stored as machine words. This cost is polynomial in $n$ for constant $k$, and it grows as $n^{O(\log n)}$ for the circuits of logarithmic depth of Sec.~\ref{sec:restart}.

\section{Numerical simulations}

In this section, we test the protocol on four families of states whose stabilizer fidelity is known exactly, and we analyze its success probability, its number of measurements and the role of the depth of the circuit.

\subsection{Simulation setup}
\label{sec:setup}

The first family consists of stabilizer states with a random perturbation, $|\psi\rangle=\sqrt{\tau}|\eta\rangle+\sqrt{1-\tau}|\chi\rangle$, with a random stabilizer state $|\eta\rangle$ and a Haar-random $|\chi\rangle$ orthogonal to it, for which $F_{\rm STAB}=\tau$ since $\tau>\cos^2(\pi/8)$ (Appendix~\ref{app:cert}). The second and third families consist of T-doped states $C\,(T|+\rangle)^{\otimes t}\otimes|0\rangle^{\otimes(n-t)}$~\cite{Leone2021,Grewal2025}, with $F_{\rm STAB}=\cos^{2t}(\pi/8)$~\cite{Rubboli2024}, which is reached by $2^t$ stabilizer states, where $C$ is a random Clifford, whose depth grows as $O(n)$, or $\lceil\log_2n\rceil$ layers of random two-qubit Cliffords between neighboring qubits, respectively. The fourth family consists of depolarized stabilizer states $(1-p)\,C|0\rangle\!\langle 0|C^\dagger+p\,I/2^n$, with the same circuits of logarithmic depth, for which $F_{\rm STAB}=1-p+p/2^n$. For each family and parameter, we simulated 15 states of up to 14 qubits, starting always from a random stabilizer state, with $N=1000$, a limit of 200 iterations, and $N_f=10234$.

Each shot was sampled from the exact probability distribution of the outcomes of the measured Pauli string, which follows from the expectation values of its substrings (Appendix~\ref{app:sims}). For the perturbed states, these expectation values were taken from the Pauli spectrum of the state, that is, the expectation values of all its $4^n$ Pauli strings, computed once from its dense state vector, which limited these states to $n\le8$. For T-doped and depolarized stabilizer states, they were computed from the Clifford circuit for each string, without storing the state. In all cases, the store keeps the substrings of weight at most $k$, and a run succeeds when the fidelity of its output, which we compute exactly, equals $F_{\rm STAB}$.

\begin{figure*}[t]
\includegraphics[width=\textwidth]{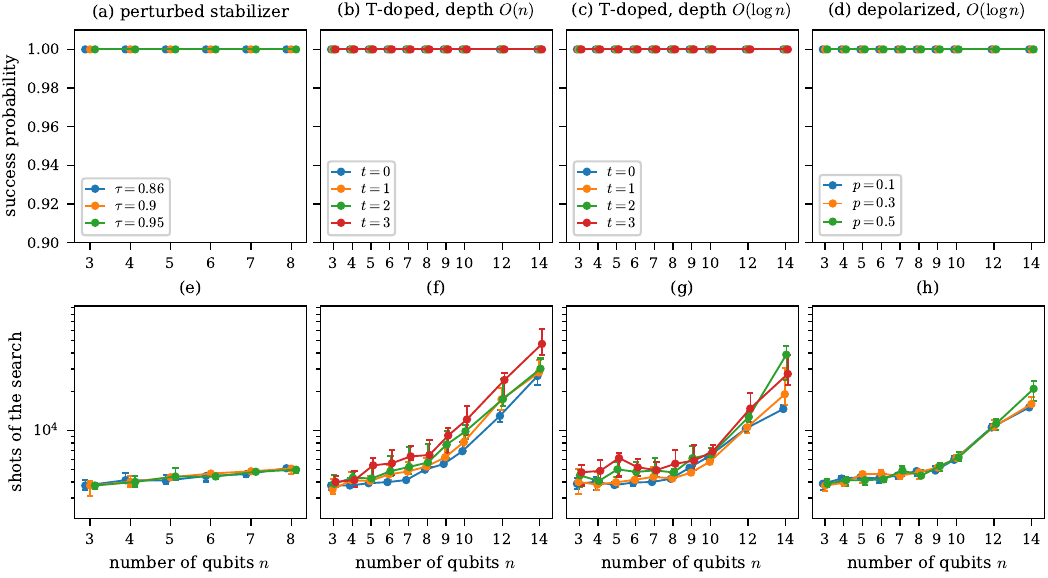}
\caption{Self-guided estimation protocol starting from a random stabilizer state. Insets (a)--(d) show the success probability, that is, the fraction of runs whose output has fidelity $F_{\rm STAB}$, as a function of the number $n$ of qubits, for (a) stabilizer states with a random perturbation of overlap $\tau$, (b) T-doped states with $t$ magic states and a random Clifford circuit, (c) T-doped states with $\lceil\log_2n\rceil$ layers of two-qubit Cliffords between neighboring qubits, and (d) depolarized stabilizer states prepared by the same circuits. Insets (e)--(h) show the median number of shots of the search, without the $N_f$ shots of the final DFE, where error bars correspond to the interquartile range. Fifteen states for each point.}
\label{fig:families}
\end{figure*}

\subsection{Success probability}

Figures\thinspace\ref{fig:families}(a)--(d) show the success probability as a function of $n$ for the four families. As is apparent from this figure, the protocol reached the optimum in all runs. Since every output is optimal and $\hat F$ is a fresh DFE of it, the error $|F_{\rm STAB}-\hat F|$ is that of a DFE of the optimal state. In addition, the certificate was obtained in all runs whose stabilizer fidelity exceeds $\cos^2(\pi/8)$ by more than $\varepsilon$, and in none of the runs with $F_{\rm STAB}\le\cos^2(\pi/8)$, where a certificate would be false. Thereby, our simulations indicate that the protocol finds the optimum of the four families from a random start.

\subsection{Number of measurements}

\begin{table}[t]
\caption{Median number of iterations $I$ of the search and median over runs of the number of shots per iteration $S/I$, in units of $10^3$ shots, as a function of the number $n$ of qubits, for $N=1000$ and all parameters of each family.}
\label{tab:resources}
\begin{ruledtabular}
\begin{tabular}{llcccccc}
 & & $n=3$ & 6 & 8 & 10 & 12 & 14 \\
\hline
Perturbed & $I$ & 3 & 3 & 3 & - & - & - \\
 & $S/I$ & 1.1 & 1.5 & 1.7 & - & - & - \\
T-doped $O(n)$ & $I$ & 4 & 4 & 3.5 & 4 & 5 & 6 \\
 & $S/I$ & 1.0 & 1.2 & 1.6 & 2.1 & 3.6 & 5.3 \\
T-doped $O(\log n)$ & $I$ & 4 & 4 & 4 & 4 & 4 & 5 \\
 & $S/I$ & 1.0 & 1.2 & 1.3 & 1.5 & 2.5 & 4.3 \\
Depolarized & $I$ & 3 & 3 & 3 & 3 & 3 & 3 \\
 & $S/I$ & 1.1 & 1.4 & 1.5 & 2.0 & 3.2 & 5.0 \\
\end{tabular}
\end{ruledtabular}
\end{table}

Figures\thinspace\ref{fig:families}(e)--(h) show the median number of shots of the search as a function of $n$. We can see that this number increases with $n$, faster for the largest simulated $n$, and that, on average over $t$, it is largest for T-doped states prepared by random Cliffords. In order to identify the origin of this growth, Table\thinspace\ref{tab:resources} separates the number of shots into the number of iterations $I$ and the number of shots per iteration $S/I$. The number of iterations remains small and increases only slightly with $n$, that is, the search reaches the optimum with a few moves, typically one or two of them given by the local restart, followed by the two iterations of the fresh check. In contrast, the number of shots per iteration increases with $n$, since the fraction of stored strings in a group of $2^n$ elements decreases (Sec.~\ref{sec:store}), so that the DFE of each new candidate samples more strings according to Eq.\thinspace\eqref{eq:store}. Therefore, the growth of the number of measurements comes from the decreasing reuse of the Pauli store, and not from a longer search.

For $2^n\gg N$, the reuse is lost, and each iteration costs a few DFEs with $N$ shots, plus about $N$ shots for each level of the prefix search whose subgroup contains more than $N$ elements. Since the simulated range ends shortly after $2^n\approx N$, it does not determine the asymptotic number of shots. For circuits of logarithmic depth, however, a string of weight $k$ appears in a fraction of about $4^{-k}\ge1/(16n^4)$ of the shots, so that, in the worst case, the local restart needs a number of shots of the order of $4^k$ times the $2\ln(2M/\delta)/\langle P\rangle^2$ outcomes required by the test of each string. Thereby, the number of shots is polynomial in $n$ for these circuits, while the classical cost of Eq.\thinspace\eqref{eq:cost} grows as $n^{O(\log n)}$.

\subsection{Depth of the circuit}
\label{sec:depth}

The protocol also reached the optimum for T-doped states prepared by random Cliffords, as shown in Fig.\thinspace\ref{fig:families}(b), although their generators are not local in general. This is because $k$ is comparable with $n$ in the simulated range, so that a random stabilizer group contains many elements of weight at most $k$. However, these elements have a larger weight than for circuits of logarithmic depth, and a string of weight $w$ appears in a fraction of about $4^{-w}$ of the shots, so that few of them pass the significance test for the largest $n$, and the candidate relies also on the strings that are not significant, which increases the number of shots, as shown in Fig.\thinspace\ref{fig:families}(f). In addition, the state built by the local restart may be improvable only along directions of large weight, which the prefix search of the fresh check finds, at least for exact values (Appendix~\ref{app:walsh}). For larger $n$, the number of elements of weight at most $k$ in a random stabilizer group decreases quickly, while circuits of logarithmic depth keep generators of small weight. All of this indicates that the depth of the circuit appears first in the cost of the protocol and, for larger $n$, in the success of the local restart.

\section{Discussion}

Estimating the stabilizer fidelity requires an optimization over $2^{O(n^2)}$ stabilizer states, which is difficult even when the fidelity of each candidate can be estimated efficiently. In this article, we proposed a self-guided estimation protocol that performs this optimization with Pauli measurements on single copies. The protocol obtains the fidelities of the stabilizer basis of the current state from a single DFE, which identifies the few directions that can improve, it reuses the outcomes of previous measurements through a Pauli store, and it builds from the few-qubit expectation values of the store a restart point that mitigates the barren plateau of a random start.

Our simulations of up to 14 qubits showed that the protocol reached the optimum in all runs for perturbed stabilizer states, T-doped states, and depolarized stabilizer states, without false certificates. The median number of iterations remained small and increased only slightly with $n$, so that the number of measurements grew mainly through the decreasing reuse of the Pauli store. For comparison, an exhaustive search requires $2^{O(n^2)}$ classical operations, even though classical shadows with random Clifford measurements estimate all the fidelities from $O(n^2/\varepsilon^2)$ copies~\cite{Huang2020}.

The local restart mitigates the barren plateau because the generators of small weight have expectation values of order one, while the fidelities from a random start are exponentially small. The cost of this strategy is set by the weight $k$ of the generators, since a string of weight $k$ appears in a fraction of about $4^{-k}$ of the shots, which gives a number of shots polynomial in $n$ and a classical cost $n^{O(\log n)}$ for circuits of logarithmic depth. For deep circuits, however, a random stabilizer group has almost no element of small weight as $n$ grows, so that in this regime the search must start close to the optimum.

The methods with theoretical guarantees for the stabilizer fidelity use Bell measurements on two copies of the state~\cite{Gross2021,Grewal2023,Arunachalam2024,Bao2024,Iyer2024,Chen2024} or adaptive Clifford measurements on single copies~\cite{Bittel2026}. They succeed for deep circuits, since Bell difference sampling, or the computational difference sampling of the adaptive scheme, reveals the stabilizer group of the optimum independently of depth, at the expense of entangling gates between two copies or between qubits in one copy. In contrast, our protocol only rotates and measures single qubits, and it provides a heuristic estimate below the fidelity thresholds of Refs.~\cite{Grewal2023,Bittel2026}, at the expense of requiring shallow circuits or a good starting point. States prepared by geometrically local shallow circuits can also be learned from single-qubit measurements~\cite{Huang2024}, but computing $F_{\rm STAB}$ from the learned description requires again the maximization of Eq.\thinspace\eqref{eq:fstab}. Furthermore, unlike the local cost functions of variational algorithms~\cite{Cerezo2021BP,Zambrano2023}, our protocol searches over a discrete set of states and obtains the local information from the data of the search itself, and, unlike the discrete search over Clifford circuits of Ref.\,\cite{Ravi2022}, whose objective is evaluated classically, it is guided by measurements of the unknown state.

Our protocol finds direct application in certifying how close the states prepared by shallow circuits in current devices are to a stabilizer state, such as the noisy code states of error-correcting codes~\cite{Gottesman1997}. In contrast, the magic states of distillation protocols~\cite{BravyiKitaev2005} have a stabilizer fidelity of at most $\cos^2(\pi/8)$, where the protocol only provides a lower bound. The main limitation of the protocol is the memory of the Pauli store, which keeps $\binom{w}{\le k}$ substrings per shot, and approaches based on classical shadows~\cite{Huang2020}, which keep only the measured strings and their outcomes, could reduce it. In addition, the use of the shots could be refined, for instance by scoring the selected candidate on fresh shots before moving, or by measuring each level of the prefix search only until its comparisons with the threshold are decided. Finally, a recent scheme bounds the fidelities of different non-stabilizer targets from the same local Pauli measurements~\cite{Li2026}, so that it could be combined with the Pauli store to bound the fidelities of the candidates without additional measurements.

\begin{acknowledgments}
L.P. was supported by the Government of Spain (Severo Ochoa CEX2019-000910-S, FUNQIP, and QEC4QEA PCI2025-163167), European Union (PASQuanS2.1, 101113690 and QEC4QEA, 101194322), Fundaci\'o Cellex, Fundaci\'o Mir-Puig, and Generalitat de Catalunya (CERCA program).
J. G. acknowledges the support of the European Union under grant agreement No. 101194322 (QEC4QEA).
\end{acknowledgments}

\appendix

\section{Properties of the stabilizer neighbors}
\label{app:neighbors}

In this appendix, we derive the fidelity of the neighbors, the directions that can improve on the current state, the Pauli strings shared by the neighbors, and the uniqueness of the optimum above $\cos^2(\pi/8)$.

\subsection{Stabilizer basis}
Let us consider a stabilizer state $|\phi\rangle$ with generators $s_ig_i$, that is, $g_i|\phi\rangle=s_i|\phi\rangle$, and destabilizers $d_i$, which commute among themselves and anticommute only with $g_i$~\cite{Aaronson2004}. Then, the product $D_x=\prod_i d_i^{x_i}$ and the elements $s_ug_u=\prod_i(s_ig_i)^{u_i}$ of $S$ satisfy
\begin{equation}
D_x\,s_ug_u\,D_x=(-1)^{x\cdot u}\,s_ug_u ,
\label{eq:conj}
\end{equation}
since $x\cdot u$ counts the generators in $s_ug_u$ that anticommute with $D_x$. Therefore, Eq.\thinspace\eqref{eq:conj} gives $(-1)^{x\cdot u}s_ug_u|\phi_x\rangle=|\phi_x\rangle$ for $|\phi_x\rangle=D_x|\phi\rangle$, so that $|\phi_x\rangle$ is stabilized by the Abelian group
\begin{equation}
S_x=\big\{(-1)^{x\cdot u}s_ug_u\mid u\in\mathbb{F}_2^n\big\},
\label{eq:Sx}
\end{equation}
which is generated by $(-1)^{x_i}s_ig_i$. That is, $S_x$ contains the same Pauli strings as $S$, with the signs of the generators with $x_i=1$ flipped. Therefore, for $x\neq y$ there is a generator $(-1)^{x_i}s_ig_i$ with opposite eigenvalues on $|\phi_x\rangle$ and $|\phi_y\rangle$, so that these states are orthogonal, and, since $D_x$ is unitary, the $2^n$ states $|\phi_x\rangle$ form an orthonormal basis. 

\subsection{Fidelity of the neighbors and directions that can improve}

In this subsection, we bound the fidelity of the neighbors of $|\phi\rangle$ in each direction $|\phi_x\rangle$, showing that only a few directions can improve the current fidelity.

The neighbors are the superpositions $|\nu_{x,m}\rangle=(|\phi\rangle+i^m|\phi_x\rangle)/\sqrt2$ of Eq.\thinspace\eqref{eq:neighbors}, with overlap $1/2$ with $|\phi\rangle$~\cite{Garcia2017}. In the basis $\{|\phi_x\rangle\}$, the values $p_x=\langle\phi_x|\rho|\phi_x\rangle$ are the populations of $\rho$, and $r_x=\langle\phi|\rho|\phi_x\rangle$ is the coherence between $|\phi\rangle$ and $|\phi_x\rangle$. Expanding the fidelity of a neighbor in this basis gives the first expression of Eq.\thinspace\eqref{eq:neighborfid}. Since $\rho$ is positive, $r_x$ is the inner product of $\sqrt\rho|\phi\rangle$ and $\sqrt\rho|\phi_x\rangle$, and the Cauchy-Schwarz inequality gives $|r_x|\le\sqrt{p_0\,p_x}$ for pure and mixed states. Therefore, since ${\rm Re}(i^mr_x)\le|r_x|$, every neighbor in direction $x$ satisfies
\begin{equation}
f(\nu_{x,m})\le\frac{p_0+p_x}{2}+\sqrt{p_0\,p_x}=\frac{(\sqrt{p_0}+\sqrt{p_x})^2}{2},
\label{eq:nbbound}
\end{equation}
whatever the coherence $r_x$. This equation gives a necessary, but not sufficient, condition for a direction to improve on the current fidelity. In fact, the bound exceeds $p_0$ only if $\sqrt{p_0}+\sqrt{p_x}>\sqrt{2p_0}$, that is, $p_x>(3-2\sqrt2)\,p_0$, so that the directions below this threshold cannot improve, while a direction above it improves only if its coherence $r_x$ is large enough. Since the $p_x$ are the probabilities of a measurement in the basis $\{|\phi_x\rangle\}$, they sum to one, and $\ell$ directions above this threshold require $\ell(3-2\sqrt2)\,p_0<1-p_0$. For instance, at most five directions can improve for $p_0=0.5$, independently of $n$. Conversely, the four values ${\rm Re}(i^mr_x)$ are $\pm{\rm Re}\,r_x$ and $\pm{\rm Im}\,r_x$, so that the best of the four neighbors has $f\ge(p_0+p_x)/2+|r_x|/\sqrt2$, that is, the four neighbors sample the phase of $r_x$ finely enough.

\subsection{Pauli strings shared by the neighbors}

In this subsection, we show that the DFE of a neighbor can reuse half of the outcomes of the DFE of $|\phi\rangle$. According to Eq.\thinspace\eqref{eq:Sx}, the elements with $x\cdot u=0$ have the same sign in $S$ and $S_x$, so that the subgroup $T_x=\{s_ug_u\mid x\cdot u=0\}=S\cap S_x$, of the elements of $S$ that commute with $D_x$, stabilizes both $|\phi\rangle$ and $|\phi_x\rangle$. In fact, since $[1+(-1)^{x\cdot u}]/2$ selects the elements with $x\cdot u=0$, Eqs.\thinspace\eqref{eq:projector} and \eqref{eq:conj} give its projector
\begin{equation}
\frac{1}{2^{n-1}}\sum_{x\cdot u=0}s_u\,g_u=|\phi\rangle\langle\phi|+|\phi_x\rangle\langle\phi_x| ,
\label{eq:Txproj}
\end{equation}
so that $T_x$ stabilizes exactly the plane spanned by $|\phi\rangle$ and $|\phi_x\rangle$. This plane behaves as one qubit, with $s_jg_j$ and $D_x$ as logical $Z$ and $X$ operators, where $g_j$ is any generator with $x_j=1$, and its six stabilizer states are $|\phi\rangle$, $|\phi_x\rangle$ and the four neighbors in direction $x$, which form the octahedron of Sec.~\ref{sec:neighbors}. Thereby, the stabilizer group of a neighbor contains the $2^{n-1}$ elements of $T_x$, with the same signs as in the stabilizer group of $|\phi\rangle$, and one more generator, namely the logical operator that has the neighbor as eigenstate with eigenvalue $+1$,
\begin{equation}
h_{x,m}=\begin{cases}(-1)^{m/2}\,D_x, & m\ \text{even},\\ (-1)^{(m-1)/2}\,i\,s_jD_xg_j, & m\ \text{odd},\end{cases}
\label{eq:hxm}
\end{equation}
where the factor $i$ makes $i\,s_jD_xg_j$ Hermitian, since $D_x$ and $g_j$ anticommute~\cite{Garcia2017}. The other half of the stabilizer group of the neighbor consists of the elements of $T_x$ multiplied by $h_{x,m}$, which do not belong to the stabilizer group of $|\phi\rangle$. Therefore, a DFE of a neighbor draws half of its strings from those of the DFE of $|\phi\rangle$. In addition, the neighbors with $m=0$ and $m=2$ have the same strings in the second half with opposite signs, as do those with $m=1$ and $m=3$, so that the outcomes measured for one of them also serve the other. However, this saves shots only when the strings of the second half are already stored, which requires few qubits, with $2^{n-1}\lesssim N$.

\subsection{Uniqueness of the optimum and certificate}
\label{app:cert}

In this subsection, we show that at most one stabilizer state has a fidelity above $\cos^2(\pi/8)$, and that the final DFE certifies when the output is this state. Two different stabilizer states satisfy $|\langle\eta_1|\eta_2\rangle|^2\le1/2$~\cite{Garcia2017}, so that the sum of their fidelities is bounded by the largest eigenvalue $1+|\langle\eta_1|\eta_2\rangle|\le1+1/\sqrt2$ of $|\eta_1\rangle\langle \eta_1|+|\eta_2\rangle\langle \eta_2|$, that is, $\langle \eta_1|\rho|\eta_1\rangle+\langle \eta_2|\rho|\eta_2\rangle\le2\cos^2(\pi/8)$ for any $\rho$. Therefore, a stabilizer state with fidelity above $\cos^2(\pi/8)$ is the unique optimum of Eq.\thinspace\eqref{eq:fstab}, as shown for pure states in Ref.\,\cite{Grewal2023}. In order to certify the optimum, we use that the $N_f$ terms of the final DFE are independent and lie in $[-1,1]$, so that the Hoeffding inequality gives $f(\phi)\ge\hat F-\sqrt{2\ln(1/\delta)/N_f}\ge\hat F_{\rm low}$, since $N_f=\lceil2\ln(1/\delta)/\varepsilon^2\rceil$, with probability at least $1-\delta$~\cite{Flammia2011,daSilva2011}. Thereby, if $\hat F_{\rm low}>\cos^2(\pi/8)$, the output is the global optimum with probability at least $1-\delta$, which for $\varepsilon=0.03$ requires $\hat F>0.8836$.

\section{Unbiasedness of the Pauli store}
\label{app:store}

In this appendix, we show that the estimate of Eq.\thinspace\eqref{eq:store} is unbiased for queries fixed in advance, and we explain why it is biased in the search. Let $U$ be the unsigned group of the target, with signs $s_P$, and let its fidelity be the mean $A_U$ of $s_P\mathrm{tr}(\rho P)$ over $U$. For queries fixed in advance, the reused set $K$ of Eq.\thinspace\eqref{eq:store} depends only on which strings were measured before, and not on their outcomes, and every stored outcome of $P$, as well as the parity of a substring, has mean $\mathrm{tr}(\rho P)$. Therefore, the first term of Eq.\thinspace\eqref{eq:store} is an unbiased estimate of the contribution of $K$ to $A_U$, the second one of the contribution of $R$, since its strings are drawn uniformly, and $\hat A_U$ is an unbiased estimate of $A_U$, whatever the correlations between strings that share a shot. In the search, in contrast, a candidate is selected because its estimate is large, so that the stored outcomes that selected it are biased upward for that candidate, and the search maximizes the estimate built from the stored outcomes rather than the fidelity itself. For this reason, the fresh check adds new shots before the search stops, and the reported fidelity is always estimated with fresh copies, whose estimate is unbiased for the output state.

\section{Fidelities of the stabilizer basis}
\label{app:walsh}

In this appendix, we show how a single DFE yields the fidelities of the stabilizer basis of $|\phi\rangle$, how the protocol determines the best sign flip, and how the prefix search identifies the directions above the threshold. Equations\thinspace\eqref{eq:projector} and \eqref{eq:conj} give $|\phi_x\rangle\langle \phi_x|=D_x|\phi\rangle\langle \phi|D_x$, so that, with $c(u)=s_u\mathrm{tr}(\rho\,g_u)$,
\begin{equation}
p_x=\frac{1}{2^n}\sum_u(-1)^{x\cdot u}\,c(u),
\label{eq:px}
\end{equation}
that is, $p_x$ is the Walsh transform of $c(u)$, the analog of the Fourier transform for functions of bit strings, as the error rates of a Pauli channel are the Walsh transform of its eigenvalues~\cite{Flammia2020}. For instance, $x=0$ gives Eq.\thinspace\eqref{eq:dfe}, and for one qubit with $|\phi\rangle=|0\rangle$ it gives $p_0=(1+\langle Z\rangle)/2$ and $p_1=(1-\langle Z\rangle)/2$. The DFE of $|\phi\rangle$ provides pairs $(u_j,v_j)$ with $u_j$ uniform and $v_j=s_{u_j}P_j$ of mean $c(u_j)$, so that $\hat p_x=\frac1N\sum_j(-1)^{x\cdot u_j}v_j$ is an unbiased estimate of $p_x$, with variance at most $1/N$, and with the Pauli store the same expression holds with the weights of Eq.\thinspace\eqref{eq:store}. Therefore, each $\hat p_x$ requires no additional measurements and $O(L)$ operations, where $L$ is the number of group elements with data, whereas the full transform would require $2^n$ values.

The best sign flip maximizes $\hat p_x$, a sum of $L$ signed parities, which is hard in general, but simple when one basis state carries most of the weight. If $p_{x^*}>1/2$, Eq.\thinspace\eqref{eq:px} gives $|c(u)-(-1)^{x^*\cdot u}p_{x^*}|\le1-p_{x^*}<p_{x^*}$, so that every $c(u)$ has the sign $(-1)^{x^*\cdot u}$, and $x^*$ is the unique solution of $x\cdot u=[c(u)<0]$ for any $n$ independent elements $u$. The protocol solves this system with the estimated signs, which are wrong with probability $(1-|c(u)|)/2$ for an element measured once, taking first the elements with the largest contributions to $\hat p_0$, and improves this solution and $x=0$ by a coordinate ascent over the bits of $x$, keeping the better endpoint. The second starting point is necessary, since the ascent from $x=0$ can stop when the best flip changes two or more generators, and each single-bit step can have zero gain, as in $\rho=|\phi_{x^*}\rangle\langle \phi_{x^*}|$. The flips of three or more generators are found from this starting point or by the prefix search of the fresh check.

The directions above the threshold are found bit by bit, as the large Fourier coefficients in the Goldreich-Levin and Kushilevitz-Mansour algorithms~\cite{Goldreich1989, Kushilevitz1993}, and finding the large error rates of sparse Pauli channels is a related problem~\cite{Harper2021}. The sum of $p_x$ over the directions whose first $j$ bits are $\alpha$ is $Q_j(\alpha)=\mathbb{E}_{u\in G_j}[(-1)^{\alpha\cdot u}c(u)]$, where $G_j$ is the subgroup generated by the first $j$ generators, so that $Q_j$ is given by a DFE on $G_j$. Since the $Q_j(\alpha)$ of each level sum to one, at most $1/\theta$ prefixes reach $\theta$, and the protocol keeps the prefixes with $\hat Q_j\ge\theta/2$, with $\theta=(3-2\sqrt2)\,\hat p_0$, at most $B=\min\{64,\lceil2(1-\hat p_0)/\theta\rceil+1\}$ of them, and estimates each level with the Pauli store, which already contains the new shots of the fresh check, while the threshold $\theta$ and the last level use the estimate of $|\phi\rangle$ of step (1). The cap of 64 only applies for $\hat p_0<0.16$, far from the optimum. Then, the directions with $\hat p_x\ge\theta$ are explored in decreasing order of $\hat p_x$, at most $\min\{B,\max\{1,\lfloor(1-\hat p_0)/\theta\rfloor\}\}$ of them, a number independent of $n$, since further directions above the threshold can only come from statistical noise, and the prefix search is skipped for $\hat p_0\le0.05$, where the local restart provides the moves. For exact values, this search finds every direction above the threshold, while with estimates it is limited by the statistical noise of $\hat Q_j$ and by the cap $B$. In addition, each level whose subgroup contains more than $N$ elements requires about $N$ new shots.

\section{Simulation details}
\label{app:sims}

In this appendix, we describe the simulation of the measurements, the exact values of the stabilizer fidelity, and the tests on the barren plateau. All simulations were carried out with Qiskit~\cite{Qiskit}.

The measurement of a Pauli string is simulated exactly, since the joint distribution of the $w$ outcome bits of a string of weight $w$ is the Walsh transform of the expectation values of its $2^w$ substrings, from which we sample the outcomes. For a T-doped state $C|m\rangle$, with $|m\rangle=(T|+\rangle)^{\otimes t}\otimes|0\rangle^{\otimes(n-t)}$, the expectation value of $P$ is that of the signed Pauli string $C^\dagger PC$, obtained letter by letter from the images of the single-qubit Pauli operators, on the product state $|m\rangle$, that is, a product of single-qubit factors, with $\langle X\rangle=\langle Y\rangle=1/\sqrt2$ and $\langle Z\rangle=0$ for $T|+\rangle$. For depolarized stabilizer states, the expectation values of the stabilizer state are multiplied by $1-p$, except for the identity. These expectation values agree with those of the dense state vector for $n\le6$. The circuits of logarithmic depth consist of $\lceil\log_2n\rceil$ layers of random two-qubit Cliffords on the pairs $(q,q+1)$, starting at even and odd $q$ in alternate layers.

The reference values of $F_{\rm STAB}$ are exact. For perturbed states, $F_{\rm STAB}=\tau$ according to Appendix~\ref{app:cert}. For depolarized stabilizer states, the fidelity with a stabilizer state $|\eta\rangle$ is $(1-p)|\langle \eta|C|0\rangle|^2+p/2^n$, which is largest for $|\eta\rangle=C|0\rangle$. For T-doped states, $F_{\rm STAB}$ does not change under Clifford gates nor when stabilizer qubits are attached, so that it equals the stabilizer fidelity $\cos^{2t}(\pi/8)$ of $(T|+\rangle)^{\otimes t}$~\cite{Rubboli2024}, which we confirmed for $t\le3$ by enumerating all stabilizer states of $t$ qubits~\cite{Dehaene2003}.

\begin{table}[b]
\caption{Number of runs that reached the optimum from a random start and median number of shots of the search, for 20 T-doped states with $t=2$ and $n=9$, ten of them prepared by random Cliffords and ten by $\lceil \log_2n\rceil = 4$ layers of two-qubit Cliffords between neighboring qubits, with $N=1000$. The first two rows use exact fidelities and no local restart.}
\label{tab:tests}
\begin{ruledtabular}
\begin{tabular}{lcc}
Search & Optimum & Shots$/10^3$ \\
\hline
All flips and neighbors, exact & 20/20 & - \\
Moves of step (2), exact & 0/20 & - \\
Protocol without local restart & 20/20 & 93.8 \\
Local candidate only & 12/20 & 3.0 \\
Protocol & 20/20 & 5.5 \\
\end{tabular}
\end{ruledtabular}
\end{table}

The statements of Sec.~\ref{sec:restart} on the barren plateau are based on the tests of Table\thinspace\ref{tab:tests}. Without local restart, the moves of step (2) did not reach the optimum even with exact fidelities, while a greedy search over all sign flips and neighbors with exact fidelities did. With $N=1000$ shots, the protocol without local restart also reached the optimum, since the statistical noise makes it accept moves that do not improve until it leaves the region of small fidelities, but it needed many more measurements than the protocol. Conversely, the protocol with the local candidate as its only move needed fewer measurements, but it stopped below the optimum in several runs, mostly for random Cliffords. Thereby, the local restart leads the search to the region of high fidelity, and the moves through the graph of neighbors complete it.

\section{Example of a local candidate}
\label{app:example}

\begin{figure}[t]
\includegraphics[width=\columnwidth]{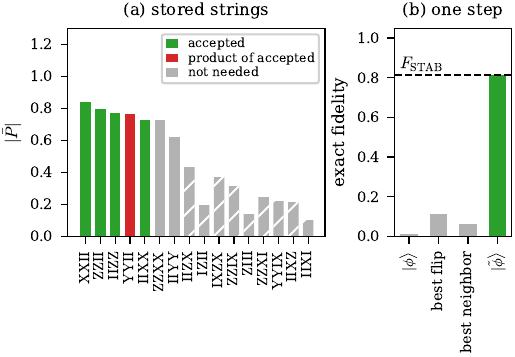}
\caption{Local candidate from the Pauli store. Inset (a) shows the stored strings in the order of the local restart, colored by the result of the greedy selection, where hatched bars are not significant at the level $\delta/M$, with $\delta=0.01$ and $M=255$. Inset (b) shows the exact fidelity of the initial state, of its best sign flip and best neighbor, and of the local candidate $|\tilde\phi\rangle$, where the dashed line marks $F_{\rm STAB}$.}
\label{fig:example}
\end{figure}

In this appendix, we follow the construction of a local candidate for four qubits in the state $\rho=0.8\,|\Phi^+\rangle\!\langle \Phi^+|^{\otimes2}+0.2\,I/16$, whose closest stabilizer state $|\Phi^+\rangle^{\otimes2}$ has generators of weight two with expectation values $\pm0.8$, and $F_{\rm STAB}=0.8125$. The search starts at a random stabilizer state with generators $-XXXI$, $-YIYX$, $-ZIYZ$ and $-YYIX$, whose fidelity is 0.0125, so that it lies in the flat region of the search. The DFE of this state with $N=1000$ shots, together with the substrings of every shot, provides outcomes for 91 Pauli strings, the first of which are shown in Table\thinspace\ref{tab:example} and in Fig.\thinspace\ref{fig:example}(a) in the order of the local restart. Exactly seven strings are significant at the level $\delta/M=0.01/255$, namely the six elements of weight two of the optimal group and $ZZXX$, while the following ones have expectation value zero.

\begin{table}[b]
\caption{Stored strings in the order of the local restart after one DFE of a random stabilizer state, for the state of Appendix~\ref{app:example}, with the number of outcomes, the mean outcome, the exact expectation value, the Hoeffding $p$-value and the result of the greedy selection. The first seven strings are significant.}
\label{tab:example}
\begin{ruledtabular}
\begin{tabular}{lccccl}
$P$ & $m_P$ & $\bar P$ & $\langle P\rangle$ & $p_P$ & Selection \\
\hline
$XXII$ & 62 & $+0.839$ & $+0.8$ & $6.8\times10^{-10}$ & accepted \\
$ZZII$ & 58 & $+0.793$ & $+0.8$ & $2.4\times10^{-8}$ & accepted \\
$IIZZ$ & 78 & $+0.769$ & $+0.8$ & $1.9\times10^{-10}$ & accepted \\
$YYII$ & 59 & $-0.763$ & $-0.8$ & $7.0\times10^{-8}$ & product \\
$IIXX$ & 58 & $+0.724$ & $+0.8$ & $5.0\times10^{-7}$ & accepted \\
$ZZXX$ & 58 & $+0.724$ & $+0.8$ & $5.0\times10^{-7}$ & not needed \\
$IIYY$ & 68 & $-0.618$ & $-0.8$ & $4.7\times10^{-6}$ & not needed \\
$IIZX$ & 60 & $+0.433$ & 0 & $7.2\times10^{-3}$ & not needed \\
\end{tabular}
\end{ruledtabular}
\end{table}

The greedy selection accepts $XXII$, $ZZII$ and $IIZZ$, rejects $YYII$ since $XXII\cdot ZZII=-YYII$, and accepts $IIXX$, which completes four generators. Then, the candidate $|\tilde\phi\rangle$ with generators $+XXII$, $+ZZII$, $+IIZZ$ and $+IIXX$ is the optimal state $|\Phi^+\rangle^{\otimes2}$. The store gives $\hat f(\tilde\phi)=0.777>\hat p_0=-0.020$, so that the search jumps from a fidelity of 0.0125 to 0.8125 in a single step, while the best sign flip and the best neighbor of the initial state have fidelities 0.1125 and 0.0625, as shown in Fig.\thinspace\ref{fig:example}(b), and a fresh DFE with $N_f=10234$ shots gives $\hat F=0.820$. Notice that in this example every stored string has tens of outcomes, so that the strings with $\langle P\rangle=0$, such as $IIZX$ with $|\bar P|=0.433$, are not significant. For larger $n$, however, most stored strings have few outcomes, and an ordering by $|\bar P|$ alone would place first many of them with $|\bar P|=1$.

\bibliography{refs}

\end{document}